# Theoretical SAXS Analysis of Structures of Disordered Gold Polymer Thin Films

M. Cattani[1] and J. M. F. Bassalo[2]

[1]Instituto de Física da Universidade de São Paulo.C.P. 66318, CEP 05315 - 970 São Paulo, SP, Brasil.  E-mail < mcattani@if.usp.br >

[2]Fundação Minerva, Av.Gov. José Malcher 629, CEP 66035 –100, Belém, PA, Brasil

E-mail <bassalo@amazon.com.br>

Abstract

Recently the structure of disordered gold-polymer(PMMA−Au) thin films have been measured using Small Angle X-ray Scattering (SAXS). We have analyzed these experimental results using the fractal SAXS theory. The films are formed of polymer−embedded gold nanoclusters and were fabricated by very low energy gold ion implantation into PMMA (polymethylmethacrylate). The composite films span (with dose variation) the transition from electrically insulating to electrically conducting regimes, a range of interest fundamentally and technologically. We find excellent agreement with theory, and show that the PMMA-Au films have monodispersive or polydispersive characteristics depending on the implanted ion dose. *Keywords:*  SAXS theory, Fractality, percolation, correlation effects.

**I. Introduction.**

Small Angle X-ray Scattering (SAXS) techniques, together with suitable theoretical analysis using models that account for disordered system peculiarities, are of fundamental importance for the characterization of nanostructured materials[1]. Aggregates of metallic nanoparticles[2,3], silica[4,5] and globular micelles[6] have been investigated in this way[7,8].

In this paper work we analyze measured SAXS scattering intensities with the appropriate theoretical models[4-9] to determine the disordered structure of PMMA-Au composite films. These kinds of composites possess interesting and technologically important electrical[10, 11, 12], lithographic[13] and optical properties.[11, 12, 14,15] The films investigated here were made by implanting gold ions at very low energy (about 49 eV) into PMMA polymer; (very low energy ion implantation is often referred to as "subplantation", and this is the appropriate terminology here). The subplanted ions form a ~6 nm wide layer of polymer-passivated gold nanoclusters lying just ~4 nm below the PMMA surface (i.e., the layer extends from ~4 nm to ~10 nm below the surface). The cluster size is about 6 nm.[16]

The formation and characteristics of gold nanoclusters are a current research issue because of their potential application in areas such as photonics and biosensing[17-22].The PMMA polymer has very good optical and dielectric properties and it immediately passivates the implanted ions. Additionally, PMMA provides a convenient route to microfabrication of the nanoparticle composite, since it can readily be lithographed using several different kinds of irradiation (electron beam, X-ray and deep-UV)[23-26]. Micro- or nanolithography of the composite allows the efficient manufacture of miniature shapes for specific applications.

We have found in SAXS a very powerful tool for investigation of our aggregates, given the variety of theoretical models available that include real system features such as fractality[27,28] and other physical correlations. The understanding of such properties is of extreme importance for understanding of the electrical, optical and chemical behavior that can arise. Thus we have utilized SAXS analysis to provide detailed information about the structural properties of our metal-polymer composites.

In Section II we show the measured[29] scattering intensities I(q) of the PMMA–Au films. In Section III the measured $I(q)$ are analyzed using appropriate theoretical approaches in order to determine the fundamental structures of the PMMA–Au films. In Section IV we discuss the results and provide a conclusion.

## II. The SAXS Scattering Intensities I(q).

The details of fabrication of the films that will be analyzed here can be seen elsewhere.[29] However, it is important to note that that the ion implantation was done using a repetitively-pulsed cathodic arc plasma gun, described in detail elsewhere[10,13,16,30–33] Three different kinds of subplanted samples were made using three different implanted gold ion doses, chosen according to the electrical characteristics[10,13] of the PMMA/gold materials that show different behavior at different doses. The conductivity mechanism for our metal-polymer composites[10] can be described in the frame of percolation theory[34].The material composed of small conducting elements in geometric contact embedded in an insulating medium can be modeled as a random resistor network and the percolation here refers to the flow of current through it[35,36]. The dc conductivity $\sigma$ of the composite, near the critical insulator/conductor transition, is given by the power law $\sigma \approx \sigma_0(x-x_c)^t$ where $\sigma_0$ is a proportionality constant, $x$ is the normalized atom concentration of the conducting phase, $x_c$ is the critical concentration, or the percolation threshold, below which the composite has zero conductivity, and $t$ is the "critical exponent".

There are three conduction regimes, depending on the implanted gold ion dose $\phi$: conductor ($\phi > \phi_c$), where $\phi_c$ is the critical dose to attain the critical concentration $x_c$, transition ($\phi \approx \phi_c$) and insulator ($\phi < \phi_c$). Our analyzed samples were formed using doses of $1.5 \times 10^{16}$ cm$^{-2}$, $1.0 \times 10^{16}$ cm$^{-2}$ and $0.8 \times 10^{16}$ cm$^{-2}$; we refer these samples as films 1, 2 and 3, respectively. Film 1 is a conducting metal-polymer composite ($\phi > \phi_c$) with gold nanoclusters that are (statistically) in geometric contact, forming percolating[10] conducting paths through the buried layer in the film. Film 2 was formed at a dose corresponding to the percolation threshold ($\phi \approx \phi_c$), and film 3 was formed at a dose less than that corresponding to the percolation threshold ($\phi < \phi_c$).

SAXS (Small Angle X-ray Scattering) measurements were carried[29] out in vacuum ambient using a SAXS Nanostar System from Bruker Instruments. In this technique as employed here[7,8], X-rays at a wavelength of 1.5418 Å are passed through a small region of the sample, approximately 3 mm$^2$, and scattered by the sample. The X-ray scattering intensity I(q) is measured as a function of the scattering vector $\boldsymbol{q}$ whose modulus is given by q = $(4\pi/\lambda)\sin(\theta/2)$, where $\lambda$ is the X-ray wavelength and $\theta$ is the scattering angle. Since our composite films are macroscopically isotropic (within the subplanted layer), the intensities depend only on the modulus of $\boldsymbol{q}$, which for SAXS is given by q $\approx (2\pi/\lambda)\,\theta$. The magnitude $q$ of $\boldsymbol{q}$ is directly linked to the size of the scattering objects; lower $q$ values correspond to larger features while higher $q$ values correspond to smaller scattering centers.

Figure 1 shows the experimentally measured intensities $I(q)$ as a function of $q$ for films 1, 2 and 3. In the next section we analyze these results using suitable SAXS theoretical models.

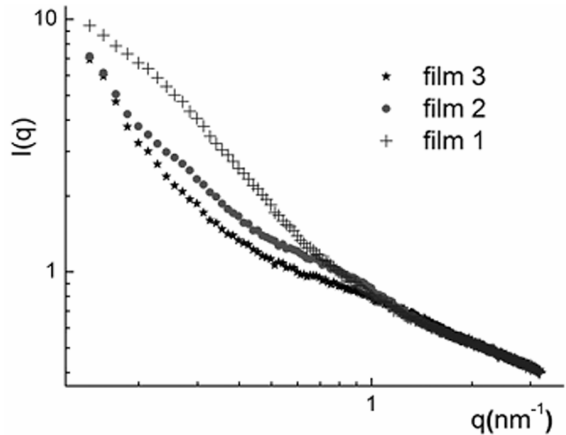

**Fig. 1** SAXS-measured $I(q)$ vs. $q$ for films 1, 2 and 3, formed with gold doses $1.5 \times 10^{16}$ cm$^{-2}$, $1.0 \times 10^{16}$ cm$^{-2}$ and $0.8 \times 10^{16}$ cm$^{-2}$, respectively.

### III. SAXS Theoretical Analysis.

To determine the fundamental structural properties of our three PMMA−Au films we analyze the experimental intensities I(q) using well established SAXS theoretical approaches [1,4-8,37]. Thus, we will analyze the $I^{(n)}(q)$ (for the films $n = 1, 2$ and 3) spectra for two different $q$ regions. In Section III.1, $I^{(n)}(q)$ will be described by a decaying exponential function $I^{(n)}(q) \sim q^{-\gamma(n)}$, as given by the "*generalized Porod law*". In Section III.2, $I^{(n)}(q)$ will be described by a *structure form factor*, and for the best fit determination of the parameters described we use the differential evolution algorithm proposed by Wormington[38].

### III.1. The Intensities $I(q) \sim q^{-\gamma}$ Described by the Generalized Porod Law.

We verify that for relatively large q values the intensities $I^{(n)}(q)$ (for the films n = 1, 2 and 3) can be fitted by a decaying exponential function, $I^{(n)}(q) \sim q^{-\gamma(n)}$. For film 1, $I^{(1)}(q) \sim q^{-0.56}$ for q in the range $\{q^{(1)}\} = 3.13 \geq q \geq 1.0$ nm$^{-1}$; for film 2, $I^{(2)}(q) \sim q^{-0.53}$ for $q$ in the range $\{q^{(2)}\} = 3.13 \geq q \geq 0.5$ nm$^{-1}$; and for film 3, $I^{(3)}(q) \sim q^{-0.57}$ for q in the range $\{q^{(3)}\} = 3.13 \geq q \geq 0.4$ nm$^{-1}$.

Note that these intensities $I^{(n)}(q)$ cannot be explained by the Porod law [3–8] $I(q) \sim q^{-4}$. To describe the intensities $I^{(n)}(q) \sim q^{-\gamma(n)}$ it is necessary to take into account simultaneously the mass fractality D of the sample and the monomer surface fractality $D_s$, where we define monomer here by the smallest scattering unit of the sample. So, taking into account the effects of both fractalities, it can be shown[1,9] that I(q) must be given by

$$I(q) \sim q^{-(2D - D_s)} \qquad (1)$$

which is the "*generalized Porod law*". In this way we see that $\gamma = 2D - D_s$.

To determine the surface fractality $D_s$ it is necessary first to evaluate the mass fractality D. This will be done in what follows by analyzing our low q range experimental data.

### III.2. Intensities Described by Structure Form Factors

In this Section we will study the intensities $I^{(n)}(q)$ for q values that are outside the regions $\{q^{(n)}\}$ considered in Section (III.1). We will show that the $I^{(n)}(q)$ for low q range can be perfectly described using the *structure form factor* approach.[4-8,37]

**a) Film 1**

The $I^{(1)}(q)$ scattering data in the q range $0.14 \leq q \leq 1.00$ nm$^{-1}$ cannot be described by a decaying exponential law. Film 1, a conducting composite, was fabricated to be just above the percolation threshold of the critical insulator-conductor transition[10]. In recent work we have shown[39] that $I^{(1)}(q)$ in this q range can be well explained assuming that the material is a monodispersive composite[5,6], that is, the material fulfills the following conditions: (a) all clusters of film 1 have approximately the same number of monomers $N$ with the same gyration radius $R_g$, (b) the clusters are formed of equal spherical gold monomers with radius $r_o$ with uniform electronic density $\rho$, (c) the monomers form dense mass fractal clusters, and (d) all clusters have the same mass fractal dimension D and consequently the same correlation distance $\xi$. The correlation distance $\xi$ represents the characteristic distance above which the mass distribution in the sample is no longer described by a fractal law[37].

Since in the considered q region the mass fractality D plays the fundamental role[1,9,37], the scattering amplitude $I^{(1)}(q)$ for our monodispersive film 1 must obey the equation[4-6,37,39]

$$I^{(1)}(q) = I_o(q)S(q) = N_{o1} N^{(1)} (\Delta\rho)^2 v_o^2 F^{(1)}(q)^2 S^{(1)}(q) \quad (2)$$

where $N^{(1)}$ is the number of clusters in film 1, $N_{01}$ is the number of monomers in each cluster, $\Delta\rho = \rho - \rho_o$ and $\rho_o$ is the uniform PMMA electronic density, $v_o$ is the monomer volume. The function $F^{(1)}(q)$, called the *single-particle form factor*, is defined by

$$F^{(1)}(q) = 3 [\sin(qr_o^{(1)}) - qr_o^{(1)}\cos(qr_o^{(1)})]/(qr_o^{(1)})^3 \quad (3)$$

and finally, the function $S^{(1)}(q)$, called the *mass fractal structure factor*, or simply *structure factor* of the particle centers, is defined by

$$S^{(1)}(q) =$$

$$1 + (1/qr_o^{(1)})^{D^{(1)}} \{D^{(1)}\Gamma(D^{(1)}-1)/[1+1/(q\xi^{(1)})^2]^{(D^{(1)}-1)/2}\} \sin[(D^{(1)}-1)\tan^{-1}(q\xi^{(1)})] \quad (4)$$

where $\Gamma(x)$ is the gamma function with argument x and $\xi^{(1)}$ is the correlation length[4-6,36].

When $q\xi^{(1)} \ll 1$ and consequently $qr_o^{(1)} \ll 1$, it can be shown[4,36] that $I_o(q) \to N^{(1)}N_{o1}(\Delta\rho)^2 v_o^2$ and

$$S^{(1)}(q) \to \Gamma(D^{(1)}+1)(\xi^{(1)}/r_o^{(1)})^{D^{(1)}} \{1 - [D^{(1)}(D^{(1)}+1)q^2\xi^{(1)2}/6]\}.$$

In this q region $I^{(1)}(q)$ becomes

$$I^{(1)}(q) = N^{(1)} N_{o1} (\Delta\rho)^2 v_o^2 \Gamma(D^{(1)}+1)(\xi^{(1)}/r_o^{(1)})^{D^{(1)}} \{1 - [D^{(1)}(D^{(1)}+1)q^2\xi^{(1)2}/6]\}$$

$$\approx N^{(1)} N_{o1}(\Delta\rho)^2 v_o^2 \, \Gamma(D^{(1)}+1)(\xi^{(1)}/r_o^{(1)})^{D^{(1)}} \exp\{-[D^{(1)}(D^{(1)}+1)q^2\xi^{(1)2}/6]\} \quad (5)$$

which has[4-6,37] a generalized Guinier–type behavior.

The *generalized gyration radius* $R_g(D,\xi)$ for spherical particles is given by

$$R_g(D^{(1)},\xi^{(1)}) = [D^{(1)}(D^{(1)}+1)/2]^{1/2}\, \xi^{(1)} \quad (6)$$

instead of $R_g = (3/5)^{1/2} r_o$, predicted by Guinier[4-6,37] neglecting the mass fractality. From Eqs. (5) and (6) we verify that for fractal systems, due to correlation effects, the monomers can be taken as assembled in clusters with radius $R_g$. In other words, $R_g \sim \xi$ defines the radius of the clusters inside of which correlation effects are important.

Equations (2)–(4) were used to fit the experimental data $I^{(1)}(q)$ shown in Fig.1 in terms of the adjustable parameters $D^{(1)}$, $\xi^{(1)}$ and $r_o^{(1)}$. The best fit values obtained for these parameters are $D^{(1)} = 1.70 \pm 0.02$, $\xi^{(1)} = 13.08 \pm 0.62$ nm, and $r_o^{(1)} = 1.64 \pm 0.04$ nm. Consequently, the gyration radius defined by Eq. (6) is equal to $R_g^{(1)} = 19.81$ nm. Fig. 2 shows the measured intensities compared with our theoretical predictions. There is excellent agreement between the experimental values and the theoretical prediction.

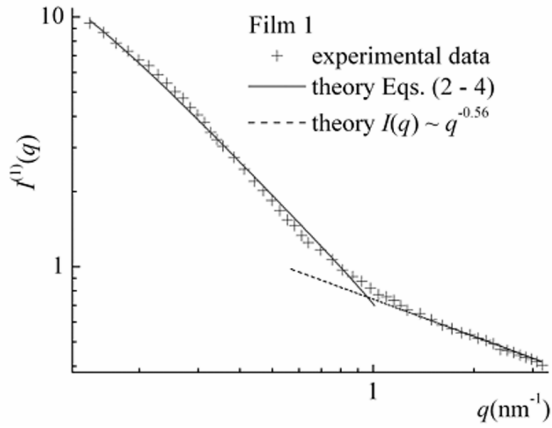

**Fig.1** $I^{(1)}(q)$ as a function q for film 1. The experimental values are indicated by crosses. The theoretical predictions given by Eqs. (2) – (4) are indicated by a continuous line in the interval $0.14 \leq q \leq 1.0$ nm$^{-1}$, and the theoretical predictions given by Eq. (1) are indicated by a dashed line in the interval $\{q^{(1)}\} = 1.00 \leq q \leq 3.13$ nm$^{-1}$.

**b) Films 2 and 3.**

For these non–percolated films 2 and 3 it was not possible to explain the intensities $I^{(2)}(q)$ and $I^{(3)}(q)$ (shown in Fig.1) assuming that they are monodispersive composites. We will show that for these films I(q) can be well explained by modeling them as *dilute polydispersive* aggregates formed by a dilute collection of $N$ fractal clusters which have radius $\{R_{gi}\}_{i=1,..,N}$, each with $N_{oi}$ monomers. Indicating the number of clusters with radius $R_{gi} = R_i$ by $N(i)_{i=1,...,N}$, the SAXS intensity generated by this polydispersive aggregate[36] is proportional to $I_{poly}(q)$ given by

$$I_{pol}(q) = \sum_{i=1..N} N(i) I_i(q) \qquad (7),$$

where

$$I_i(q) = I_{oi}(q) S_i(q),$$

$$I_{oi}(q) = N_{oi}(\Delta\rho)^2 v_o^2 P(q) = N_{oi}(\Delta\rho)^2 v_o^2 \{3[\sin(qr_o) - qr_o\cos(qr_o)]/(qr_o)^3\}^2$$

and

$$S_i(q) = 1 + (1/qr_o)^D \{D_i \Gamma(D_i-1)/[1 + 1/(q\xi_i)^2]^{(D_i-1)/2}\} \sin[(D_i-1)\tan^{-1}(q\xi_i)] \qquad (8).$$

Taking into account that the factor $(\Delta\rho)^2 v_o^2 P(q)$ is the same for all clusters, the intensity $I_{poly}(q)$ given by Eq. (2) can be written as

$$I_{poly}(q) = (\Delta\rho)^2 v_o^2 P(q) \sum_{i=1..N} N(i) N_{oi} S_i(q) \qquad (9)$$

For a *monodispersive* sample formed by $N$ identical clusters, the intensity $I_{mon}(q)$, using Eq. (9), is simply given by $I_{mon}(q) = N N_o (\Delta\rho)^2 v_o^2 P(q) S(q)$, as seen in Section (III.1).

With this model let us investigate film 2 assuming that the film is composed of only two different kind of clusters, with radius $R_{g1}^{(2)} = R_1^{(2)}$ and $R_{g2}^{(2)} = R_2^{(2)}$. Taking $R_{g2}^{(2)} > R_{g1}^{(2)}$, we have $\xi_2^{(2)} > \xi_1^{(2)}$. In this way, in the $q$ region where $q\xi_2^{(2)} \ll 1$ and $qr_o^{(2)} \ll 1$ we obtain

$$S_1^{(2)}(q) \sim \Gamma(D^{(2)}+1)(\xi_1^{(2)}/r_o^{(2)})^{D^{(2)}} \exp[-(R_1^{(2)})^2 q^2/3], \qquad (10).$$

where, $R_1^{(2)} = [D^{(2)}(D^{(2)}+1)/2]^{1/2} \xi_1^{(2)}$ according to Eqs. (5) and (6), respectively. Similarly, in the region where $q\xi_2^{(2)} \ll 1$ and $qr_o \ll 1$ we have

$$S_2^{(2)}(q) \sim \Gamma(D^{(2)}+1)(\xi_2^{(2)}/r_o^{(2)})^{D^{(2)}} \exp[-(R_2^{(2)})^2 q^2/3], \qquad (11)$$

with $R_2^{(2)} = [D^{(2)}(D^{(2)}+1)/2]^{1/2} \xi_2^{(2)}$. Consequently,

$$I^{(2)}{}_{poly}(q) \sim (\Delta\rho)^2 v_o{}^2 P(q)\{f_1{}^{(2)}S_1{}^{(2)}(q) + f_2{}^{(2)}S_2{}^{(2)}(q)\} \sim f_1{}^{(2)}I_1{}^{(2)}(q) + f_2{}^{(2)}I_2{}^{(2)}(q) \quad (12),$$
where $f_1{}^{(2)}$ and $f_2{}^{(2)}$, with $f_1{}^{(2)} + f_2{}^{(2)} = 1$, are the fractions of clusters 1 and 2, of the film 2, respectively.[36]

The function $I^{(2)}{}_{poly}(q)$ defined by Eq. (12) can be used to obtain a best fit to the experimental data $I^{(2)}(q)$ given in Fig. 3 in terms of the adjustable parameters $r_o{}^{(2)}$, $D^{(2)}$, $f_1{}^{(2)}$, $f_2{}^{(2)}$, $\xi_1{}^{(2)}$ and $\xi_2{}^{(2)}$, and these parameters can in this way be determined. With the best fit approach we found that $r_o{}^{(2)} = 1.59 \pm 0.04$, $R_{g1}{}^{(2)} = 18.80$ nm, $f_1{}^{(2)} = 0.370 \pm 0.008$, $\xi_1{}^{(2)} = 10.85 \pm 0.20$ nm, $R_{g2}{}^{(2)} = 4.34$ nm, $f_2{}^{(2)} = 0.631 \pm 0.009$ and $\xi_2{}^{(2)} = 2.98 \pm 0.08$ nm and $D^{(2)} = 1.62 \pm 0.03$.

Fig. 3 shows the measured intensities $I^{(2)}(q)$ compared to the theoretical predictions given by Eq. (12) for small q values in the range $0.14 \leq q \leq 0.41$ nm$^{-1}$, and the theoretical predictions given by $I^{(2)}(q) \sim q^{-\gamma(2)}$, where $\gamma^{(2)} = 0.53$, for q in the range $\{q^{(2)}\} = 0.41 \leq q \leq 3.13$ nm$^{-1}$. Once again we see that there is excellent agreement between theory and experiment.

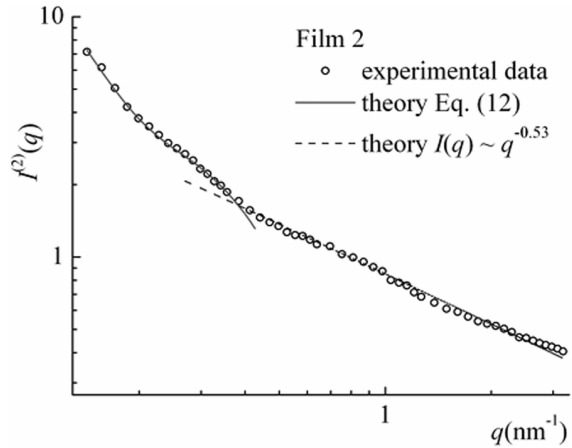

**Fig.3** $I^{(2)}(q)$ as a function of q for film 2. The experimental values are indicated by open circles. The theoretical predictions given by Eq. (12) are indicated by a continuous line in the interval $0.14 \leq q \leq 0.41$ nm$^{-1}$, and the theoretical predictions given by Eq. (1) are indicated by a dashed line in the interval $\{q^{(2)}\} = 0.41 \leq q \leq 3.13$ nm$^{-1}$.

In the case of film 3, in order to explain the intensity $I^{(3)}(q)$ we found that it was also necessary to assume that the material is polydispersive, composed also by two different kinds of clusters with radius $R_{g1}{}^{(3)}$, $R_{g2}{}^{(3)}$ associated with correlation lengths $\xi_1{}^{(3)}$, $\xi_2{}^{(3)}$, respectively. A structure function $S_n{}^{(3)}(q)$, with $n = 1$ and 2 describes each cluster. Following the same procedure adopted to analyze film 2 we now have

$$I^{(3)}_{poly}(q) \sim (\Delta\rho)^2 v_o^2 P(q)\{f_1^{(3)}S_1^{(3)}(q) + f_2^{(3)}S_2^{(3)}(q)\} \sim f_1^{(3)}I_1^{(3)}(q) + f_2^{(3)}I_2^{(3)}(q) \quad (13),$$

where $f_1^{(3)}$ and $f_2^{(3)}$, with $f_1^{(3)} + f_2^{(3)} = 1$, are the fractions of clusters 1 and 2, respectively.

The function $I^{(3)}_{poly}(q)$ defined by Eq. (13) can be used to obtain a best fit to the experimental data $I^{(3)}(q)$ shown in Fig. 1 in terms of the adjustable parameters $f_1^{(3)}$, $f_2^{(3)}$, $\xi_1^{(3)}$, $\xi_2^{(3)}$ which in this way can thus be determined. Using this best fit approach we find that $r_o^{(3)} = 1.72 \pm 0.04$, $R_{g1}^{(3)} = 14.926$ nm, $f_1^{(3)} = 0.41 \pm 0{,}01$, $\xi_1^{(3)} = 9.90 \pm 0.19$ nm, $R_{g2}^{(3)} = 5.77$ nm, $f_2^{(3)} = 0.58$, $\xi_2^{(3)} = 3.829 \pm 0.08$ nm and $D^{(3)} = 1.69 \pm 0.03$.

Fig.4 shows the measured intensities $I^{(3)}(q)$ compared to the theoretical predictions given by Eq. (13) for small q values in the range $0.14 \leq q \leq 0.41$ nm$^{-1}$, and with the theoretical predictions given by $I^{(3)}(q) \sim q^{-\gamma(3)}$, where $\gamma^{(3)} = 0.53$ for q in the range $\{q^{(3)}\} = 0.41 \leq q \leq 3.13$ nm$^{-1}$. From Fig. 4 we observe that there is a very good agreement between theory and experiment.

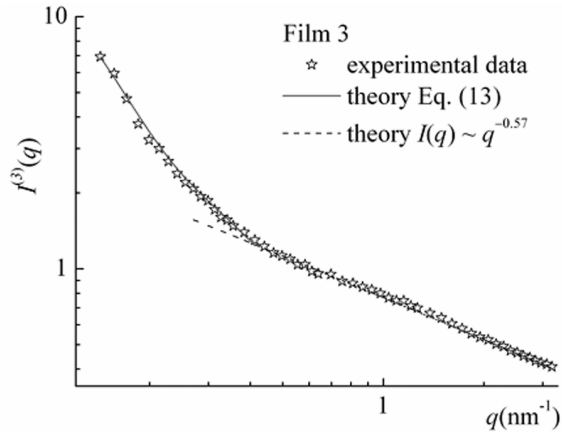

**Fig.4**. $I_3(q)$ as a function q for film 3. The experimental values are indicated by open stars. The theoretical predictions of Eq. (13) are indicated by a continuous line in the interval $0.14 \leq q \leq 0.41$ nm$^{-1}$, and the theoretical predictions given by Eq. (1) are indicated by a dashed line in the interval $0.41 \leq q \leq 3.13$ nm$^{-1}$.

Extensive studies[28,40,41] performed to determine the surface fractalities $D_s$ have shown that $3.0 \geq D_s \geq 2.0$ according to the predictions of Sayles and Thomas [39,40]. Thus, supposing that $I(q) \sim q^{-\gamma}$ are due to mass and surface monomers fractalities we must expect that $\gamma = 2D - D_s$. In this way, the fractality $D_s$ will be evaluated taking from $D_s = 2D - \gamma$ and using the known $\gamma$ and $D$ values determined in the preceding Sections. These $D_s$ values, together with the values of $\gamma$ and $D$, are shown in Table I, where we can see that they are very similar.

**Table I.** Fundamental structure parameters D, $D_s$, and $\gamma$, for films 1, 2 and 3.

| | Generalized Porod Law | | |
|---|---|---|---|
| Sample | $\gamma$ | $D_s$ | $D$ |
| Film 1 | 0.56 ± 0.01 | 2.84 | 1.70 ± 0.02 |
| Film 2 | 0.53 ± 0.01 | 2.71 | 1.62 ± 0.03 |
| Film 3 | 0.57 ± 0.02 | 2.81 | 1.69 ± 0.03 |

In Table II are seen the fundamental structure parameters: monomer sizes $r_o$, correlation lengths $\xi$, gyration radii $R_g$ and their respective fractions f.

**Table II.** Fundamental structure parameters $r_o$, $R_g$, and fraction f of the gyration radius for films 1, 2 and 3.

| | Structure form factors approach | | | |
|---|---|---|---|---|
| Sample | $R_g$ (nm) | fraction $f$ | $\xi$ (nm) | $r_o$ (nm) |
| Film 1 (monodispersive) | $R_g$ = 19.81 | ---- | $\xi$ = 13.08 ± 0.62 | 1.64 ± 0.04 |
| Film 2 (polydispersive) | $R_{g1}^{(2)}$ = 15.80  $R_{g2}^{(2)}$ = 4.34 | $f_1^{(2)}$ = 0.370 ± 0.008  $f_2^{(2)}$ = 0.631 ± 0.009 | $\xi_1^{(2)}$ = 10.85 ± 0.20  $\xi_2^{(2)}$ = 2.98 ± 0.08 | 1.59 ± 0.04 |
| Film 3 (polydispersive) | $R_{g1}^{(3)}$ = 14.92  $R_{g2}^{(3)}$ = 5.77 | $f_1^{(3)}$ = 0.41 ± 0,01  $f_2^{(3)}$ = 0.58 ± 0.01 | $\xi_1^{(3)}$ = 9.90 ± 0.19  $\xi_2^{(3)}$ = 3.82 ± 0.08 | 1.72 ± 0.04 |

## IV. Discussions and Conclusions.

The analysis described in Section 3 provides very good agreement between the theoretical predictions and the experimental results for the intensities $I^{(n)}(q)$ summarized in Section II. We have shown that for the three films, due to both monomer surface fractalities and sample mass fractalities, the scattering intensities can be explained by the generalized Porod law in the q regions defined by $\{q^{(1)}\} = 3.13 \geq q \geq 1.0$ nm$^{-1}$, $\{q^{(2)}\} = 3.13 \geq q \geq 0.5$ nm$^{-1}$ and $\{q^{(3)}\} = 3.13 \geq q \geq 0.4$ nm$^{-1}$. For relatively small q values the intensities $I^{(n)}(q)$ can be very well explained assuming film 1 to be monodispersive and films 2 and 3 polydispersive in the context of the structure form factor approach.[4-8,37]

Besides the good agreement obtained between the measured and predicted intensities $I^{(n)}(q)$, we see that our predicted values D, $D_s$, f, $R_g$, $\xi$ and $r_o$ are reasonable from a physical point of view and similar to those found in other recent work on metal-polymer composites[2,3] using SAXS.

Finally, let us remark that films 3, 2 and 1 were fabricated with increasing gold ion dose: $0.8 \times 10^{16}$ cm$^{-2}$, $1.0 \times 10^{16}$ cm$^{-2}$ and $1.5 \times 10^{16}$ cm$^{-2}$, respectively. Film 1 is a conducting metal–polymer composite and films 2 and 3 are insulators. According to our theoretical analysis we see that as the dose increases the polydispersivity decreases: the insulating film 3 and film 2 are formed by two different gyration radii and film 1, which is conducting, is monodispersive. Also, interestingly, we note an increase of the higher gyration radius as the gold dose is increased. However, the low q ranges behavior shows that the films have strong similarities concerning the mass fractal dimension and surface fractal dimension.

**Acknowledgements.**

This work was supported by the Fundação de Amparo a Pesquisa do Estado de São Paulo (FAPESP). We thank Ms. Virginia de Paiva for the support at the IFUSP library.